\documentclass[letterpaper]{article} 
\usepackage{aaai25}  
\usepackage{times}  
\usepackage{helvet}  
\usepackage{courier}  
\usepackage[hyphens]{url}  
\usepackage{graphicx} 
\usepackage{natbib}  
\usepackage{caption} 
\usepackage{algorithm}
\usepackage{algorithmicx}

\usepackage{newfloat}
\usepackage{listings}

\usepackage{multirow}
\usepackage{array}
\usepackage{colortbl}
\usepackage{booktabs}
\usepackage{hhline}
\usepackage{subcaption}
\usepackage{amssymb}
\usepackage{amsmath}
\usepackage{threeparttable}
\usepackage{mathrsfs}
\usepackage{verbatim}
\usepackage{algpseudocode} 
\usepackage{tikz} 
\usepackage{etoolbox}
\usepackage{marvosym}

\pubnote{Published as a conference paper at AAAI 2025}

\newrobustcmd*\circled[1]{\tikz[baseline=(char.base)]{
            \node[shape=circle,draw,inner sep=1pt,fill,text=white,minimum size=1em] (char) {\textsf{\small #1}};}}

\DeclareCaptionStyle{ruled}{labelfont=normalfont,labelsep=colon,strut=off} 
\floatstyle{ruled}
\newfloat{listing}{tb}{lst}{}
\floatname{listing}{Listing}
\title{Pushing the Limits of BFP on Narrow Precision LLM Inference}
\author{%
      Hui Wang$^{1 \ast}$,
      Yuan Cheng$^{2,3 \ast \dag}$,
      Xiaomeng Han$^{1}$,
      Zhengpeng Zhao$^4$,
      Dawei Yang$^{2 \textrm{\Letter}}$,
      Zhe Jiang$^{1 \textrm{\Letter}}$,
      }
\affiliations {
    \textsuperscript{\rm 1}National Center of Technology Innovation for EDA, School of Integrated Circuits, Southeast University\\
    \textsuperscript{\rm 2}Houmo AI\\
    \textsuperscript{\rm 3}Nanjing University\\
    \textsuperscript{\rm 4}Huazhong University of Science and Technology\\
    whmio0115@seu.edu.cn, yuancheng@smail.nju.edu.cn, mingzhihan7@gmail.com,
    u202114911@hust.edu.cn, 
    dawei.yang@houmo.ai, 
    zhejiang.uk@gmail.com
}

\begin{document}

\maketitle

\def\thefootnote{$\textrm{\Letter}$}\footnotetext{Corresponding authors.}
\def\thefootnote{$\ast$}\footnotetext{Equal contribution.}
\def\thefootnote{$\dag$}\footnotetext{This work was conducted during his internship at Houmo AI.}

\begin{abstract}

The substantial computational and memory demands of Large Language Models (LLMs) hinder their deployment. Block Floating Point (BFP) has proven effective in accelerating linear operations, a cornerstone of LLM workloads. 
However, as sequence lengths grow, nonlinear operations, such as Attention, increasingly become performance bottlenecks due to their quadratic computational complexity. These nonlinear operations are predominantly executed using inefficient floating-point formats, which renders the system challenging to optimize software efficiency and hardware overhead. In this paper, we delve into the limitations and potential of applying BFP to nonlinear operations. Given our findings, we introduce a hardware-software co-design framework (DB-Attn), including: (i) DBFP, an advanced BFP version, overcomes nonlinear operation challenges with a pivot-focus strategy for diverse data and an adaptive grouping strategy for flexible exponent sharing. (ii) DH-LUT, a novel lookup table algorithm dedicated to accelerating nonlinear operations with DBFP format. (iii) An RTL-level DBFP-based engine is implemented to support DB-Attn, applicable to FPGA and ASIC. Results show that DB-Attn provides significant performance improvements with negligible accuracy loss, achieving 74\% GPU speedup on Softmax of LLaMA and 10x low-overhead performance improvement over SOTA designs.

\end{abstract}

\section{Introduction}

\par The noteworthy success of Large Language Models (LLMs) has revolutionized various fields of artificial intelligence. The emerging LLMs like LLaMA families \cite{meta2024llama3} and Mistral families  \cite{jiang2023mistral} continue to push the boundaries of what these models can achieve, promising even greater capabilities in natural language understanding, generation, and problem-solving across diverse domains. 
While LLMs have exhibited remarkable performance across a range of tasks, their inference process demands substantial computing power and memory bandwidth, severely hindering their application and implementation.  

Various approaches have been explored for efficient large model deployment. While quantization \cite{lin2024qserve, ma2024affinequant} and pruning \cite{frantar2023sparsegpt, xia2023flashllmenablingcosteffectivehighlyefficient} reduce model size and complexity, they often suffer from accuracy degradation and complex post-training processes. Alternative numerical formats like BF16 and TF32 \cite{burgess2019bfloat16, choquette20213} improve efficiency but remain costly for large-scale inference. Block Floating-Point (BFP) \cite{drumond2018training, darvish2020pushing} offers a promising solution by sharing exponents within data blocks, providing the dynamic range for DNN inference with minimal hardware overhead.

\begin{figure}[t]
\centering
\includegraphics[width=0.40\textwidth]{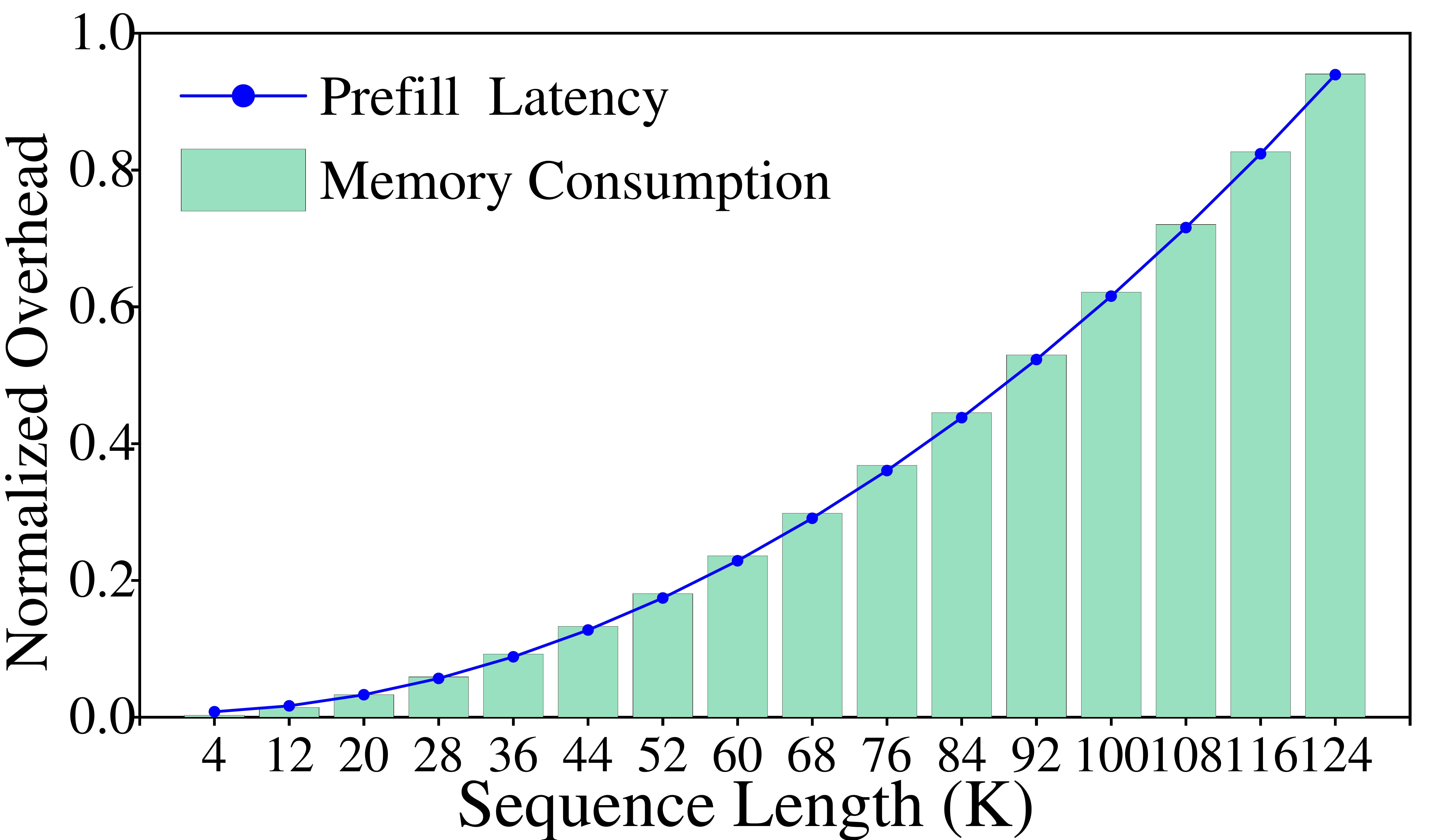}
\caption{Memory overhead and latency of prefill stages for LLaMA3-8B scale superlinearly with sequence length. 
}
\label{fig:fig1}
\end{figure}

While BFP research in deep learning has primarily targeted linear operations, such as convolution and fully connected layers, nonlinear operations like Softmax and GELU still depend on floating-point computations, emerging as performance bottlenecks.
For instance, Softmax alone consumes over 30\% of LLM inference time \cite{stevens2021softermax}. Analysis of LLaMA3-8B shows that longer sequences lead to super-linear growth in memory and latency (Fig.~\ref{fig:fig1}), due to the quadratic complexity of Attention layers.


In this paper, we pioneer the application of BFP to nonlinear operations and identify three key challenges:
\circled{1} Outlier Sensitivity: accuracy degradation from exponent alignment with outliers.
\circled{2} Representation Granularity: uniform alignment of exponent drops the representation granularity, which leads to less accurate results.
\circled{3} Hardware Complexity: nonlinear operations involve complex logic (transcendental functions, division, etc.), complicating system optimization (detailed in Tab. \ref{table:table1}).
To this end, we propose:

\noindent \textbf{Dynamic-BFP (DBFP)}, an advanced variant of BFP, adeptly addresses the challenges of accuracy and efficiency in nonlinear operations. It incorporates two novel strategies: a pivot-focus strategy capable of conforming to various data distributions, and an adaptive grouping strategy that enables a more precise and flexible exponent sharing mechanism.

\noindent \textbf{DB-Attn}, 
a DBFP-based framework with software-hardware co-optimization for efficient Attention computation. It accelerates Softmax without floating-point operations and streamlines the dataflow between linear (BFP Matmul) and nonlinear (DBFP Softmax) operations by sharing exponents, eliminating explicit conversions.

\begin{itemize}
\item \texttt{Algorithm}. We present DH-LUT, a nonlinear operations dedicated lookup table (LUT) algorithm with DBFP format, achieving 74\% speedup on the GPU for Softmax while maintaining comparable accuracy to floating-point.
\item \texttt{Hardware}. We design and implement an RTL-level DBFP-based engine applicable to FPGA and ASIC, delivering 10x throughput over SOTA designs.
\end{itemize}

\section{Related Work}

\subsection{Data formats for LLMs}

As LLMs grow in size and complexity, the standard 32-bit floating-point (FP32) format becomes less practical. 
Researchers develop low-bit formats such as BF16 \cite{jouppi2020domain} and TF32 \cite{NVIDIA2022TF32} to address increasing memory and computational demands.
Low-bit fixed-point data types convert operations to integer operations (e.g., INT4) by fixing the number of floating-point bits \cite{nagel2019data, NVIDIA2022INT4, mellempudi2017mixed}, risking accuracy drop with large dynamic ranges.
BFP formats \cite{drumond2018training} share exponents within data blocks, enabling efficient GEMM operations through dense integer logic with reduced hardware complexity. While \cite{song2018computation} explores various BFP grouping methods, they are limited by spatial constraints and ignore data distribution characteristics. Although BSFP \cite{lo2023block} improves upon BFP, its complex hardware design presents implementation challenges. The MXFP format \cite{rouhani2023microscaling} represents another step towards efficient floating-point computations in deep learning.


\subsection{Nonlinear Operation Algorithms}
Nonlinear operations present unique challenges in efficient temporal memory utilization and computation, requiring multiple passes over input values held in memory \cite{kim2023stackoptimizationtransformerinference}. Calculation methods for these operations typically fall into two categories: LUT-based method \cite{zhang2023high, zhang2022base, du2019efficient} and approximation algorithms \cite{kim2021bert, xia2023softmax, wang2018high}. 
While LUTs offer high accuracy, they demand substantial storage. Approximation, on the other hand, generally improves in accuracy with larger computational units.

\section{Methodology}

\subsection{Dynamic Block Floating-Point}

Here, we introduce the DBFP, including its mathematical model and the corresponding optimization.

\subsubsection{BFP Formulation.}
Let $\mathbb{F}$ denotes the set of floating-point numbers, and $x \in \mathbb{F}$ is represented as:
$x = (-1)^s  \cdot 2^e \cdot m$
where
$s \in \left \{ 0, 1 \right \} $ is the sign bit, 
$e \in [e_\text{min}, e_\text{max}]$ is the exponent, 
and $m \in [1, 2)$ is the mantissa.
BFP numbers are formally defined as a group:
$Y = (y_1, y_2, \ldots, y_n)$, where each  $y_i$ shares an exponent $e_{s}$,
such that $y_i = (-1)^{s}_i \cdot m_i \cdot 2^{e_\text{s}}$. 
BFP numbers can be partitioned into two fields: a shared field $S$ and a private field $P$. Each $\hat{s_j} \in S$ represents a shared exponent for multiple elements in $P$. Each $p_i \in P$ is of the form $p_i = (-1)^{s}_i \cdot m_i$. 

To convert floating-point numbers to BFP, a uniform exponential alignment is applied as Eq.\ref{eq:eq1}:

\begin{small}\begin{equation}
\begin{aligned} \label{eq:eq1}
x_i 
= (-1)^s_i  \cdot 2^{e_i} \cdot m_i 
&= (-1)^s_i  \cdot 2^{e_i-d_{ij}}\cdot (m_i \cdot 2^{d_{ij}}) \\
&= (-1)^s_i  \cdot 2^{\hat{s_j}} \cdot m_i'
\end{aligned}\end{equation}
\end{small}

, where $d_{ij} = e_i - \hat{s_j}$ denotes the difference between the exponent of $x_i$ and the $j$th shared exponent. 
This alignment introduces an error due to the finite precision of mantissa multiplication by $2^{d_{ij}}$ (shifting $d_{ij}$ bits).
The error depends on the distance $d_{ij}$, determined by the input exponent $e_i$. Therefore, selecting appropriate shared exponents $\hat{s_j}$, which are determined by many factors, is crucial to minimize $d_{ij}$.

\noindent \textbf{\underline{Problem Formulation}}:
To this end, we formulate the problem as finding a adaptive grouping function $f$ that partitions a set $X$ of floating-point values into $k$ subsets based on their characteristics:
$f: X \rightarrow \chi \left \{ S_1, S_2, ..., S_k \right \}  $
where each subset $S_j$ is determined within a frame of discernment $\Omega$.


Each subset $S_j$ has a unique BFP representation $B_j$ with shared and private fields. We then explore factors influencing BFP format accuracy, which define the function $f$.

\subsubsection{Observations and Insights.} 
Through experimental and theoretical analysis, we explore the limitations of vanilla BFP, providing new insights for improving accuracy. 

\noindent \textbf{\underline{Observation 1 Pivot-focus policy}}: Vanilla BFP formats typically align a group of $x_i$ (e.g., every 16 elements) to the maximum exponent within that group, leading to significant accuracy drop. 
Our experimental findings in Softmax show that setting the alignment direction to the default maximum causes up to 9.6x greater loss than the median one. 

\noindent \textbf{Insight 1}: 
This suggests that using maximum values for alignment is suboptimal. Instead, a more representative value (e.g., median) as an alignment pivot better preserves accuracy across the group.


\noindent \textbf{\underline{Observation 2 Adaptive grouping strategy}}: Prior work on BFP relied on fixed bounding boxes, which are particularly vulnerable to outliers.
Outliers can cause disproportionate shifts in data exponents and lead to substantial accuracy drop. See Tab. \ref{table:table1} for a more detailed analysis. 

\noindent \textbf{Insight 2}: If elements with similar magnitude distributions can share exponents within a group, it can reduce the bit shifts (diminish the $d_{ij}$) for individual numbers. 

\begin{figure}[t]
\centering
\includegraphics[width=0.47\textwidth]{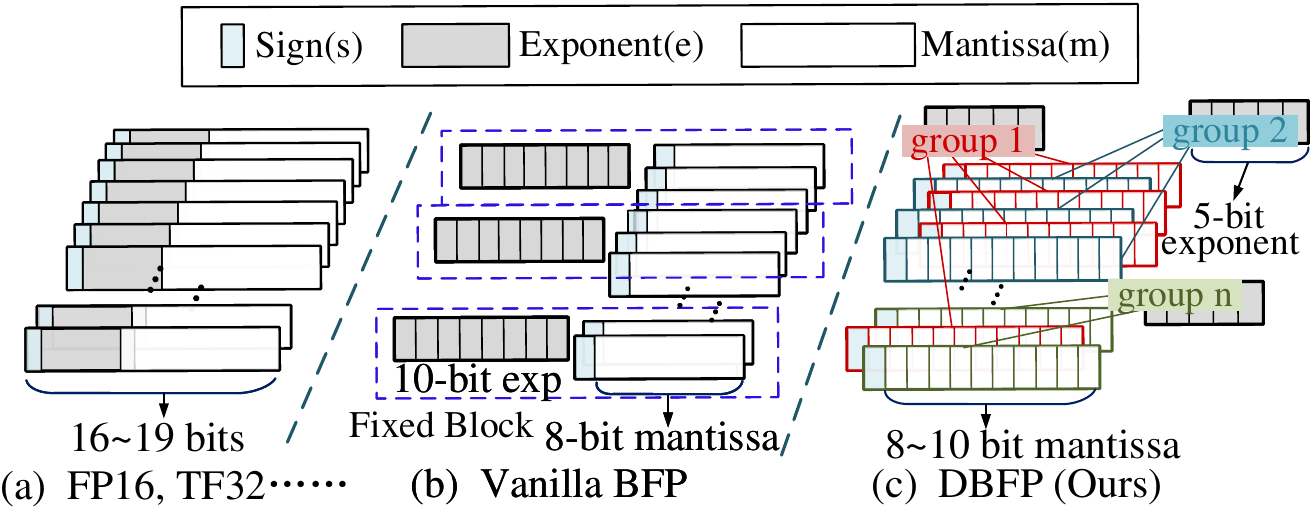}
\caption{Number system comparison between floating-point numbers(a), BFP(b) and our DBFP(c).}
\label{fig:fig2}
\end{figure}

Fig.~\ref{fig:fig2} illustrates the difference between the mentioned formats:
(a) represents floating-point data formats such as FP16 or TF32;
(b) shows the vanilla BFP format;
(c) highlights our DBFP, which employs automatic grouping based on the intervals in which the exponent falls.

\subsubsection{Theoretical Analysis.}
For the pivot-focus policy, we introduce directional vectors to characterize the process. 
Let $\vec{s_j}$ denote the vector representing the direction and magnitude from the origin to the point $\hat{s_j}$ in the vector space.
The shift vector $\vec{v_{ij}}= (\left| \vec{s_j} \right| - e_i)\hat{v}$ determines the shift for each $x_i$ relative to the sharing exponent. 
$\vec{m_{i}'}$ and $\vec{m_{i}}$ denote the mantissa's projection before and after conversion, with their Euclidean distance as 
$d_{ij}^{2} = \left\| \vec{m_{i}}'-\vec{m_{i}} \right\|^2$. $\mu_{ij}$ 
represents the confidence that $x_i$ falls within the interval defined by $\vec{s_j}$.
It's aimed to minimize the following objective function:

\begin{small}
\begin{equation} 
\label{eq:eq2}
\begin{aligned}
\mathop{\arg\min}\limits_{\mu_{ij}, \vec{s_j}} \mathcal{J}_{\mathrm{DBFP}}
&\triangleq 
\sum_{i=1}^{n} \sum_{S_{j} \subseteq \Omega} \mu_{ij}^{\beta} D_{ij}^2 \\
&= \sum_{i=1}^{n} \sum_{\left\{j / S_{j} \neq \emptyset, S_{j} \subseteq \Omega\right\}}  \mu_{ij}^{\beta} d^2_{ij}+\sum_{i=1}^{n} \mu_{i \emptyset}^{\beta} \delta^{2} 
\end{aligned}
\end{equation}
\end{small}

$s.t.$
\begin{small}
$ \displaystyle 
 \sum_{\left\{j / S_{j} \subseteq \Omega S_{j} \neq \emptyset \right\}}  \mu_{ij} +  \mu_{i \emptyset} = 1, {\forall} i=1,n, 
$
\end{small}


In Eq. \ref{eq:eq2}, hyperparameter $\beta$ (default 2) regulates $\mu_{ij}$'s importance. An empty set mitigates outlier impact on $\vec{s_j}$ selection, with $\mu_{i \emptyset}$ as its confidence. The distance between outliers and $\hat{s_{\emptyset}}$ is $\delta^{2}$, related to the distances of all $S_j$.

\begin{small}
\begin{equation}
\label{eq:eq3}
D^2_{ij}=
\left\{   
\begin{aligned}
&\delta^{2}, S_{j} = \emptyset\\
&d_{ij}^{2}, \left | S_{j} \right | = 1  \\
\end{aligned}
\right.
\end{equation}
\end{small}

To minimize $\mathcal{J}_{\mathrm{DBFP}}$, we alternately fix one of the variables $\mu_{ij}$ or $\vec{s_j}$ and solve the constrained minimization problem for the other. By introducing $n$ Lagrange multipliers $\lambda_i$, the Lagrangian is expressed as:

\begin{small}
\begin{equation}
\begin{aligned}
\label{eq:eq4}
\mathcal{L}(U, S, \lambda_{1},  \cdots  ,\lambda_{n}) = \mathcal{J}_{\mathrm{DBFP}}(U, S) - \sum_{i=1}^{n} \sum_{S_{j} \subseteq \Omega} \lambda_{i} \mu_{ij}^{\beta}  \\
\end{aligned}
\end{equation}
\end{small}

\begin{figure}
\centering
\includegraphics[width=0.47\textwidth]{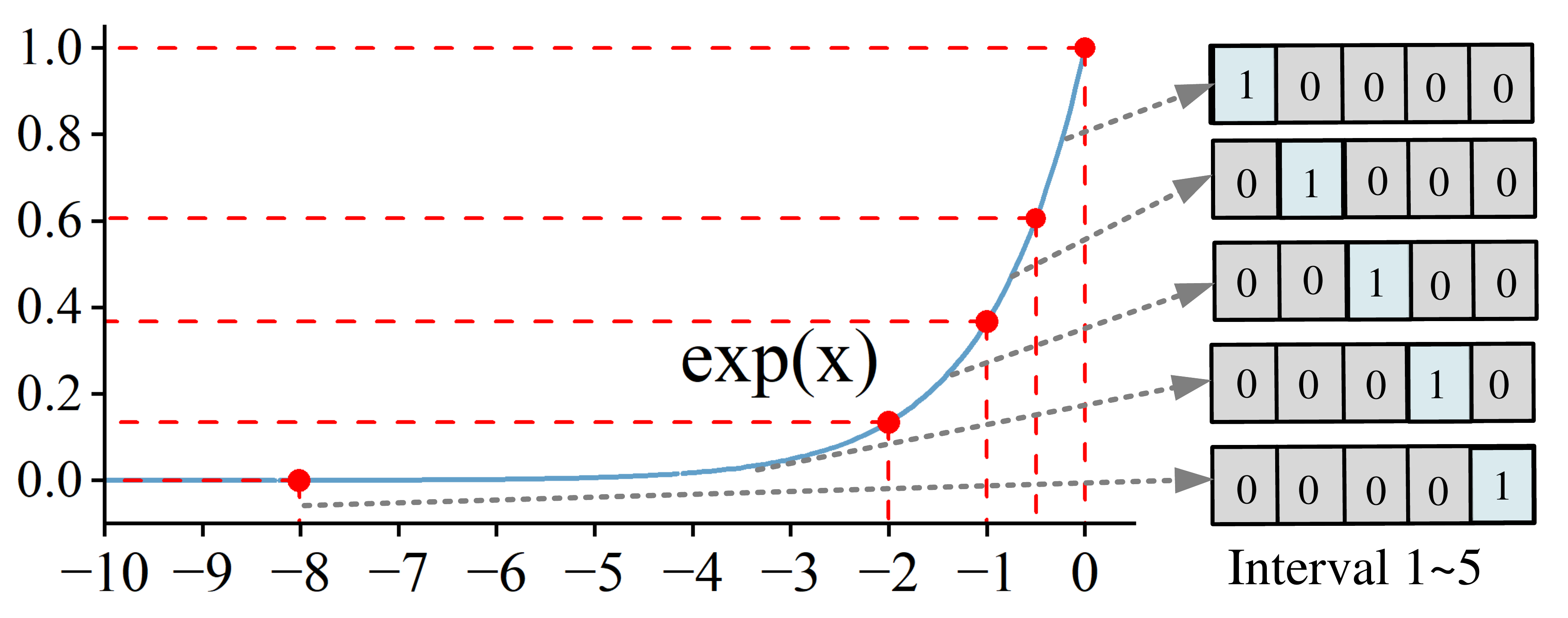}
\caption{Non-uniform hierarchical LUT with five intervals.}
\label{fig:fig3}
\end{figure}

By differentiating the Lagrangian and setting each element of gradient $\nabla \mathcal{L}$ to zero, specifically 
$
\frac{\partial \mathcal{L}}{\partial \mu_{ij}} ,
\frac{\partial \mathcal{L}}{\partial \mu_{i\emptyset}}, 
\frac{\partial \mathcal{L}}{\partial \lambda_{i}}
$
, we obtain $\mu_{ij}$, the necessary conditions for optimality:

\begin{small}
\begin{equation}
\begin{aligned}
\label{eq:eq5}
\mu_{ij} = \frac{d_{ij}^{-2/(\beta-1) }}{  {\textstyle \sum_{S_k \ne \emptyset }} d_{ik}^{-2/(\beta-1)} + \delta ^{-2/(\beta-1)}} 
\end{aligned}
\end{equation}
\end{small}

Similarly, we can further calculate the $\vec{s_{j}}$

\begin{small}
\begin{equation}
\begin{aligned}
\label{eq:eq6}
\vec{s_{j}} =  \frac{ {\sum^n_{i=1} \sum_{S_j \ne \emptyset} \mu_{ij}^\beta \vec{e_i}}} {\sum^n_{i=1} \sum_{S_j \ne \emptyset} \mu_{ij}^\beta} 
\end{aligned}
\end{equation}
\end{small}

This system can be solved using a standard linear system solver. 
Ultimately, this process yields an optimal solution for the set $S$, i.e., the configuration that complies with the adaptive grouping strategy and pivot-focus policy.

\subsection{DB-Attn Algorithm Design}
In this section, we introduce the algorithmic core of DB-Attn. Unlike the vanilla BFP format, which struggles with nonlinear operations, our proposed DH-LUT, for the first time, completes nonlinear operations in a BFP-like format. 
We also apply DBFP to linear operations, optimizing Matmul efficiency through cascade operations.

\subsubsection{Optimization for Softmax.}
The Softmax operation is among the most computationally intensive nonlinear components in Transformers, significantly impacting computational efficiency. Our optimization methodologies using DBFP offer a general approach to efficiently computing nonlinear functions, readily extensible to other nonlinear operations.
The Softmax function converts Attention scores into a probability distribution. It can be represented as follows:

\begin{small}
\begin{equation}\label{eq:eq7}
\operatorname{Softmax}\left(x_i\right)=\frac{e^{x_i}}{\sum_{j=0}^{i} e^{x_j}} = \frac{e^{x_i - x_{\text{max}}}}{\sum_{j=0}^{i} e^{x_j - x_{\text{max}}}}
\end{equation}
\end{small}

, which is transformed by the down-scaling exponentiation to smooth data distribution and prevent overflow.

Softmax's bottleneck stems from the low throughput of exponential functions. Leveraging DBFP's shared exponents, we propose Dynamic Hierarchical LUT (DH-LUT), a lightweight solution. DH-LUT uses a two-dimensional structure of sub-LUTs, dynamically loaded based on DBFP shared exponents and high $k$ bits of mantissa.
For Softmax's $e^{x - x_{\text{max}}}$ operation, only the $[-\infty, 0]$ range needs fitting, showing non-uniform distribution. DH-LUT adapts to this through pivot-focus policy and non-uniform fitting. We optimize accuracy and memory with a non-uniform hierarchical allocation method (Algo.\ref{alg:alg_1}), adaptively partitioning to focus resources on nonlinear regions shown in Fig.~\ref{fig:fig3}.

For the Softmax based on DH-LUT, its computational error stems from the DBFP format and the mapping error of DH-LUT. Eq. \ref{eq:eq8} introduces the error analysis of vanilla BFP, which is applicable to DBFP. For a BFP block, using the rounding-to-nearest scheme, its quantization error is zero-mean with variance $\sigma^{2}$, defined as:

\begin{small}
\begin{equation}\label{eq:eq8}
\sigma^{2}=\frac{2^{-2 L_{m}}}{12} \sum_{i=1}^{N_{\gamma}} p_{\gamma_{i}} 2^{2 \gamma_{i}}
\end{equation}
\end{small}

, where $L_{m}$  denote the bit length of the block mantissa, and $p_{\gamma_{i}}$ represent the probability mass function (PMF) of the block exponent. $N_{\gamma} = 2^{L_{E}}$ is the number of available block exponential levels.  
Increasing the bit length of the block mantissa and reducing the shared probability of larger exponents can effectively reduce input value errors.
While aligning to the maximum exponent (vanilla BFP) causes significant errors as inputs with tiny exponents are crucial for Softmax shown in Fig.\ref{fig:fig3}, and minimum exponent alignment preserves accuracy at the cost of redundant mantissa bit, as the table width of DH-LUT is fixed at the Pareto frontiers, detailed in Fig.\ref{fig:fig5}, our approach aligns with the median exponent, balancing accuracy and efficiency.

The data stored in DH-LUT is all DBFP, and its characteristic of shared exponents allows calculations not only in the floating-point domain but also on computing resources-limited devices, supporting integer-only computation:

\begin{equation}
{ \rm Softmax(}x_i\rm ) = \frac{e^{x_i-x_\text{max}}_\text{ DH-LUT}}{ \scriptstyle\sum^d_{j}{e^{x_j-x_\text{max}}_{\text{DH-LUT}}}} = \frac{2^\text{s} \cdot e^\text{int}_i}{ \scriptstyle \sum^d_{j}{2^\text{s} \cdot e^\text{int}_j}} = \frac{e^\text{int}_i}{ \scriptstyle \sum^d_{j}{e^\text{int}_j}}
\end{equation} \label{eq:eq9}

DB-Attn leverages the DH-LUT with minimal storage overhead to compute the exponential function, enhancing the throughput of nonlinear operations. Its shared exponent feature facilitates the migration of the Softmax algorithm to devices with limited computing resources, achieving significant speedups on both floating-point platforms and edge devices with integer-only capabilities.

\subsubsection{Optimization of Matrices.}

From Eq. \ref{eq:eq1}, a vector in DBFP format can be viewed as the product of a shared coefficient and an integer vector, expressed as: $\overrightarrow{A_{D}} = 2^{e_A} \cdot \overrightarrow{A_{I}}$
where $\overrightarrow{A_{I}}$ represents an integer vector. $e_A$ is the exponent shared by this vector.
Applying DBFP to matrices in Attention, the dot product operation of a single vector $\overrightarrow{Q_\text{D}}$ and $\overrightarrow{K^T_\text{D}}$ within the Matmul of Query and Key can be described as :

\begin{small}
\begin{equation}\label{eq:eq10}
\begin{split}
    \overrightarrow{Q_\text{D}}\cdot \overrightarrow{K^T_\text{D}} &= (2^{e_Q} \cdot \overrightarrow{Q_{I}})\cdot(2^{e_K} \cdot \overrightarrow{K^T_{I}})\\
    &= 2^{e_Q + e_K}(\overrightarrow{Q_{I}}\cdot\overrightarrow{K^T_{I}})
\end{split}
\end{equation}
\end{small}

\begin{figure*}[t]
\includegraphics[width=1.0\textwidth]{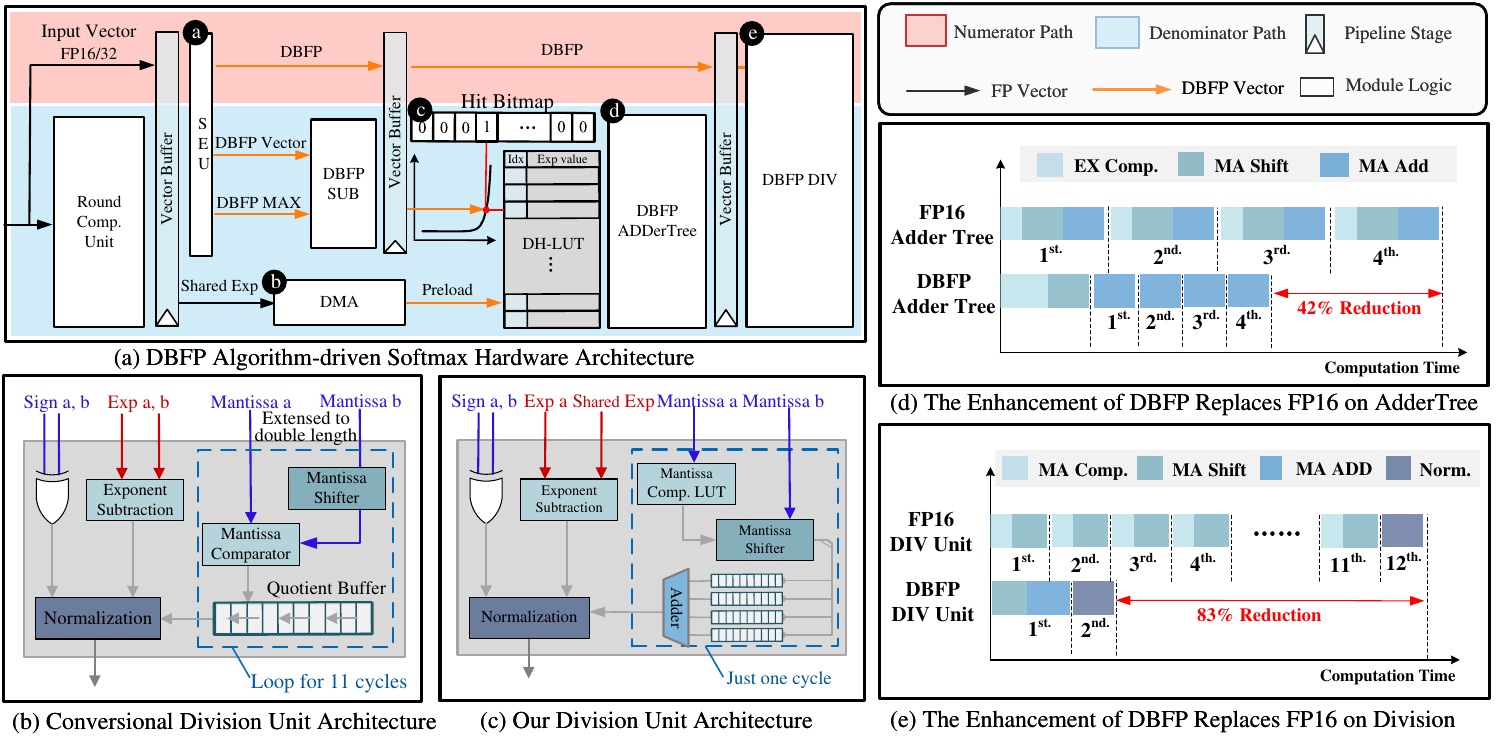}
\caption{DB-Attn algorithm-driven Softmax hardware architecture and enhancement compared with FP16 design.} \label{fig:fig4}
\end{figure*}

\begin{algorithm}[ht]
\caption{DH-LUT hierarchical algorithm}
\label{alg_lut}
\small
\textbf{Input}: Target Function $F$, lut table size $m$, all possible values under the FP16 format $V$. \\
\textbf{Output}: Optimal Partition Points of the LUT $OPP$ . 

\begin{algorithmic}[1]
\Function {Update\_next} {$OPP$, $i$, $m$, $V$}
\State $ interval = (len(x)-OPP[i])//(m - 1 - i) $ 
\For {j = 1 to m - 1 - i }
    \State $OPP[i+j] = OPP[i]+j*interval$
\EndFor
\EndFunction
\State
\Function{Select\_best\_OPP} {$F$, $V$, $m$}
\State $interval = len(x) // (m - 1)$
\State Initialize $OPP$ with $interval$
\For {i = 1 to m}
    \State $D_{max} = \infty$
    \State $pre = OPP[i-1]$
    \State $next = OPP[i+1]$
    \State $out_{F} = F(V)$
    \For {j = pre + 1 to next}
        \State calculate Interpolation $out_{right}$
        \State calculate MAE $D_{right}$  
        \State calculate Interpolation $out_{left}$
        \State calculate MAE $D_{left}$   
        \State $D = D_{left} + D_{right}$
        \State Update $OPP[i], D_{max}$ if $D \leq D_{max}$
        \State {$\Call{Update\_next}{OPP, i, m, V}$}
    \EndFor
\EndFor
\State \Return {$OPP$}
\EndFunction
\end{algorithmic}
\label{alg:alg_1}
\end{algorithm}

It can be discerned that vector dot products and Matmul in the DBFP format can be achieved using integer Matmul units and integer adders, bypassing complex floating-point operations and enhancing computational efficiency. Extending this operation to the entire matrix reveals the trait of cascaded DBFP Matmul. The result of multiplying two DBFP matrices remains a DBFP matrix, capable of being seamlessly chained for subsequent Matmul without additional handling. For DBFP matrices with finer-grained blocks, re-alignment can be introduced during cascading operations, efficiently handled via hardware shift operations, enabling even more efficient computation.

\subsection{ Algorithm-driven Hardware Architecture} 
We design and implement an RTL-level engine for Softmax with DBFP and evaluate the hardware resources and throughput advantage. 
Benefiting from the DBFP, the proposed accelerator offers competitive performance with very light design complexity, making it easily adaptable to other general-purpose GPUs and NPUs. Below, We'll first detail the proposed architecture and implementation.

\subsubsection{Overall Architecture.}
With the alignment to the algorithms presented above, we split the architecture of the accelerator into four pipeline stages:
\textbf{\texttt{Max}} finds the maximum value within the input vector; \textbf{\texttt{SE}} performs Shared Exponent calculation and maximum value subtraction; \textbf{\texttt{Exp}} computes exponents and sum using an adder tree; \textbf{\texttt{Div}} executes division operations to obtain the result.

The \texttt{SE} stage 
segments and aligns the exponent of the input vector with the \texttt{MAX} stage's maximum value, followed by subtraction (Fig.~\ref{fig:fig4} \circled{a}). 
Simultaneously, it checks if the DH-LUT's current data exponent matches the required to-be-shared exponent, signalling 
DMA (Direct Memory Access, a module allowing hardware subsystems to access memory for efficient data transfer between memory and devices) to preload new data required for the following computations (Fig.~\ref{fig:fig4} \circled{b}).
These initial stages prepare for computation. 
The subsequent three phases complete the exponential function (Fig.~\ref{fig:fig4} \circled{c}), denominator summation (Fig.~\ref{fig:fig4} \circled{d}),  and division (Fig.~\ref{fig:fig4} \circled{e}) -- the main operations in Softmax.

\subsubsection{Low-bit Storage Implementation of DBFP.}

Our dynamic hierarchical non-uniform LUT strategy introduced in the previous section enables a compact yet flexible hardware implementation. 
By extracting $n$ bits from vector elements, we create a $2^n$-entry table that balances accuracy and size. 
This compact design is suitable for Softmax computation, which requires global input information.
Our approach utilizes two tables: a value table storing exp values for approximation and a hit bitmap table recording mantissa occurrences (Fig.~ \ref{fig:fig4} \circled{c}). 
Input vectors perform lookups in parallel, with each exponent index hitting a DH-LUT interval, setting a corresponding bitmap bit. 
After recording the input vector, we multiply and sum values from both tables using an adder tree structure. 
This achieves parallel lookup results without extra hardware resources, leveraging the compact table size.

\begin{table*}[t]
  \centering

    \resizebox{.95\textwidth}{!}{
    \begin{tabular}{cccccccccccc}
    \toprule
    \multirow{2}[4]{*}{Model} & \multirow{2}[4]{*}{Method} & \multirow{2}[4]{*}{Nonlinear Op} & \multicolumn{7}{c}{Zero-Shot}                         & \multicolumn{2}{c}{Perplexity} \\
\cmidrule(lr){4-10}\cmidrule(lr){11-12}   \cmidrule(lr){4-10}       &       &       & PIQA(↑)\newline{} & ARC-e(↑) & ARC-c(↑) & BoolQ(↑) & HellaSwag(↑) & Winogrande(↑)\newline{} & \multicolumn{1}{p{4.19em}}{avg.(↑)} & WikiText2(↓) & C4(↓) \\
    \midrule
    \multirow{6}[2]{*}{LLaMA-7b} & FP16  & FP32  & 77.37  & 52.27  & 41.21  & 73.27  & 72.87  & 67.32  & 64.05  & 5.68  & 7.08  \\
          & 
          \cellcolor[rgb]{ .906,  .902,  .902}BFP   & 
          \cellcolor[rgb]{ .906,  .902,  .902}BFP   & 
          \cellcolor[rgb]{ .906,  .902,  .902}53.65  & 
          \cellcolor[rgb]{ .906,  .902,  .902}30.05  & 
          \cellcolor[rgb]{ .906,  .902,  .902}25.34  & 
          \cellcolor[rgb]{ .906,  .902,  .902}61.87  & 
          \cellcolor[rgb]{ .906,  .902,  .902}32.06  & 
          \cellcolor[rgb]{ .906,  .902,  .902}49.41  & 
          \cellcolor[rgb]{ .906,  .902,  .902}42.06  & 
          \cellcolor[rgb]{ .906,  .902,  .902}67.31  & 
          \cellcolor[rgb]{ .906,  .902,  .902}67.13  \\
          & FP8 e4m3 & FP32  & 49.51  & 25.08  & 22.69  & 37.82  & 25.04  & 49.57  & 34.95  & NaN   & NaN \\
          & FP8 e4m3-S & FP32  & 49.46  & 25.17  & 22.78  & 37.82  & 25.04  & 49.57  & 34.97  & NaN   & NaN \\
          & FP8 e5m2 & FP32  & 77.31  & \textbf{51.98}  & 41.04  & 72.14  & 72.24  & 65.66  & 63.40  & 5.80  & 7.16  \\
          & \cellcolor[rgb]{ .906,  .902,  .902}\textbf{DBFP} & \cellcolor[rgb]{ .906,  .902,  .902}\textbf{DH-LUT} & \cellcolor[rgb]{ .906,  .902,  .902}\textbf{77.64}  & \cellcolor[rgb]{ .906,  .902,  .902}51.85  & \cellcolor[rgb]{ .906,  .902,  .902}\textbf{41.47}  & \cellcolor[rgb]{ .906,  .902,  .902}\textbf{73.30}  & \cellcolor[rgb]{ .906,  .902,  .902}\textbf{72.87}  & \cellcolor[rgb]{ .906,  .902,  .902}\textbf{67.01}  & \cellcolor[rgb]{ .906,  .902,  .902}\textbf{64.02}  & \cellcolor[rgb]{ .906,  .902,  .902}\textbf{5.68}  & \cellcolor[rgb]{ .906,  .902,  .902}\textbf{7.08}  \\
    \midrule
    \multirow{6}[2]{*}{LLaMA2-7b} & FP16  & FP32  & 76.88  & 53.62  & 40.61  & 71.07  & 72.94  & 67.09  & 63.70  & 5.47  & 6.97  \\
          & \cellcolor[rgb]{ .906,  .902,  .902}BFP   
          & \cellcolor[rgb]{ .906,  .902,  .902}BFP   
          & \cellcolor[rgb]{ .906,  .902,  .902}52.72  
          & \cellcolor[rgb]{ .906,  .902,  .902}28.24  
          & \cellcolor[rgb]{ .906,  .902,  .902}25.26  
          & \cellcolor[rgb]{ .906,  .902,  .902}61.90  
          & \cellcolor[rgb]{ .906,  .902,  .902}33.64  
          & \cellcolor[rgb]{ .906,  .902,  .902}49.41  
          & \cellcolor[rgb]{ .906,  .902,  .902}41.86  
          & \cellcolor[rgb]{ .906,  .902,  .902}32.72  
          & \cellcolor[rgb]{ .906,  .902,  .902}41.29  \\
          & FP8 e4m3 & FP32  & 49.51  & 25.08  & 22.69  & 37.82  & 25.04  & 49.57  & 34.95  & NaN   & NaN \\
          & FP8 e4m3-S & FP32  & 49.56  & 25.08  & 22.69  & 37.82  & 25.04  & 49.57  & 34.96  & NaN   & NaN \\
          & FP8 e5m2 & FP32  & \textbf{76.98}  & 52.35  & 40.69  & 70.55  & 72.09  & 65.43  & 63.02  & 5.61  & 7.09  \\
          & \cellcolor[rgb]{ .906,  .902,  .902}\textbf{DBFP} & \cellcolor[rgb]{ .906,  .902,  .902}\textbf{DH-LUT} & \cellcolor[rgb]{ .906,  .902,  .902}76.77  & \cellcolor[rgb]{ .906,  .902,  .902}\textbf{53.20}  & \cellcolor[rgb]{ .906,  .902,  .902}\textbf{41.47}  & \cellcolor[rgb]{ .906,  .902,  .902}\textbf{71.01}  & \cellcolor[rgb]{ .906,  .902,  .902}\textbf{72.60}  & \cellcolor[rgb]{ .906,  .902,  .902}\textbf{67.32}  & \cellcolor[rgb]{ .906,  .902,  .902}\textbf{63.73}  & \cellcolor[rgb]{ .906,  .902,  .902}\textbf{5.48}  & \cellcolor[rgb]{ .906,  .902,  .902}\textbf{6.98}  \\
    \midrule
    \multirow{6}[2]{*}{LLaMA3-8b} & FP16  & FP32  & 80.79  & 77.74  & 53.33  & 81.35  & 79.17  & 72.61  & 74.17  & 6.14  & 8.88  \\
          & \cellcolor[rgb]{ .906,  .902,  .902}BFP   
          & \cellcolor[rgb]{ .906,  .902,  .902}BFP   
          & \cellcolor[rgb]{ .906,  .902,  .902}54.30  
          & \cellcolor[rgb]{ .906,  .902,  .902}31.06  
          & \cellcolor[rgb]{ .906,  .902,  .902}22.78 
          & \cellcolor[rgb]{ .906,  .902,  .902}49.17  
          & \cellcolor[rgb]{ .906,  .902,  .902}35.67  
          & \cellcolor[rgb]{ .906,  .902,  .902}49.72  
          & \cellcolor[rgb]{ .906,  .902,  .902}40.45  
          & \cellcolor[rgb]{ .906,  .902,  .902}69.65  
          & \cellcolor[rgb]{ .906,  .902,  .902}79.49  \\
          & FP8 e4m3 & FP32  & 49.51  & 25.08  & 22.70  & 37.82  & 25.04  & 49.56  & 34.95  & NaN   & NaN \\
          & FP8 e4m3-S & FP32  & 49.73  & 25.17  & 22.61  & 37.82  & 25.04  & 49.56  & 34.99  & NaN   & NaN \\
          & FP8 e5m2 & FP32  & 79.59  & 78.32  & 52.55  & 79.41  & 78.05  & 71.50  & 73.24  & 6.42  & 9.24  \\
          & \cellcolor[rgb]{ .906,  .902,  .902}\textbf{DBFP} & \cellcolor[rgb]{ .906,  .902,  .902}\textbf{DH-LUT} & \cellcolor[rgb]{ .906,  .902,  .902}\textbf{80.36}  & \cellcolor[rgb]{ .906,  .902,  .902}\textbf{78.66}  & \cellcolor[rgb]{ .906,  .902,  .902}\textbf{53.16}  & \cellcolor[rgb]{ .906,  .902,  .902}\textbf{81.01}  & \cellcolor[rgb]{ .906,  .902,  .902}\textbf{78.86}  & \cellcolor[rgb]{ .906,  .902,  .902}\textbf{73.48}  & \cellcolor[rgb]{ .906,  .902,  .902}\textbf{74.26}  & \cellcolor[rgb]{ .906,  .902,  .902}\textbf{6.14}  & \cellcolor[rgb]{ .906,  .902,  .902}\textbf{8.89}  \\
    \bottomrule
    \end{tabular}%
    }
      \caption{Accuracy performance of DB-Attn in different LLM tasks}\label{table:table1}
\end{table*}%

\begin{table*}[t]
  \centering
  \resizebox{.95\textwidth}{!}{
    \begin{tabular}{cccccccccccc}
    \toprule
    Model & Method & Nonlinear Op & Map   & Model & Method & Nonlinear Op & Top-1 acc. & Model & Method & Nonlinear Op & Top-1 acc. \\
    \midrule
    \multirow{6}[2]{*}{DETR} & FP32  & FP32  & 41.9  & \multirow{6}[2]{*}{ViT-base} & FP32  & FP32  & 84.536 & \multirow{6}[2]{*}{Swin-tiny} & FP32  & FP32  & 81.378 \\
          & \cellcolor[rgb]{ .906,  .902,  .902}BFP   & \cellcolor[rgb]{ .906,  .902,  .902}BFP   & \cellcolor[rgb]{ .906,  .902,  .902}26.8  &       & \cellcolor[rgb]{ .906,  .902,  .902}BFP   & \cellcolor[rgb]{ .906,  .902,  .902}BFP   & \cellcolor[rgb]{ .906,  .902,  .902}39.132 &       & \cellcolor[rgb]{ .906,  .902,  .902}BFP   & \cellcolor[rgb]{ .906,  .902,  .902}BFP   & \cellcolor[rgb]{ .906,  .902,  .902}72.052 \\
          & FP8 e4m3 & FP32  & NaN   &       & FP8 e4m3 & FP32  & 84.482 &       & FP8 e4m3 & FP32  & 81.312 \\
          & FP8 e4m3-S & FP32  & NaN   &       & FP8 e4m3-S & FP32  & 84.442 &       & FP8 e4m3-S & FP32  & 81.400 \\
          & FP8 e5m2 & FP32  & 28.4  &       & FP8 e5m2 & FP32  & 84.246 &       & FP8 e5m2 & FP32  & 81.268 \\
          & \cellcolor[rgb]{ .906,  .902,  .902}\textbf{DBFP} & \cellcolor[rgb]{ .906,  .902,  .902}\textbf{DH-LUT} & \cellcolor[rgb]{ .906,  .902,  .902}\textbf{41.8} &       & \cellcolor[rgb]{ .906,  .902,  .902}\textbf{DBFP} & \cellcolor[rgb]{ .906,  .902,  .902}\textbf{DH-LUT} & \cellcolor[rgb]{ .906,  .902,  .902}\textbf{84.522} &       & \cellcolor[rgb]{ .906,  .902,  .902}\textbf{DBFP} & \cellcolor[rgb]{ .906,  .902,  .902}\textbf{DH-LUT} & \cellcolor[rgb]{ .906,  .902,  .902}\textbf{81.384} \\
    \bottomrule
    \end{tabular}%
    }
  \caption{Accuracy performance of DB-Attn in different Vision tasks}
    \label{table:table2}
\end{table*}%

\subsubsection{Efficient DBFP Computing.}
DBFP's ability to convert floating-point operations into integer operations is a key advantage. 
By sharing exponents, most floating-point computations simplify basic exponent operations combined with integer mantissa operations, streamlining arithmetic calculations.
Vector addition in neural networks exemplifies this efficiency. Traditional floating-point addition requires aligning exponents for each number pair. DB-Attn pre-aligns exponents within groups, reducing the operation to exponential multiplication with mantissa addition.
For cascading structures like adder trees (Fig.~\ref{fig:fig4} \circled{d}), a single initial exponent alignment allows direct mantissa calculations, yielding substantial benefits. Fig.~\ref{fig:fig4}(d) demonstrates an experimentally proven 42\% latency reduction in a 4-level adder tree. 

Division operations in Softmax and Layernorm often limit parallelism. DBFP's shared exponents enable efficient parallel integer division approximation (Fig.\ref{fig:fig4} \circled{e}), reducing latency of these complex operations. While traditional FP16 dividers require 11 cycles for 10-bit mantissa calculation, consuming 90\% of area and power, our DBFP Divider uses LUTs and shift-addition \cite{jha2020fpad} to complete division in a single cycle. In Softmax, with fixed divisor and identical DBFP exponents, we need only one exponent subtraction and lookup per 64 divisions, achieving 83\% latency reduction (Fig.\ref{fig:fig4}(e)). This approach enables FP16 operations using 10-bit integer operations, particularly benefiting bit-width sensitive operations.



\section{Evaluation}

\noindent \textbf{Baselines.}
As described, DB-Attn involves both software and hardware implementation. 
Hence, we examined DB-Attn against the SOTA work in both worlds.
For the software, we compare DB-Attn with FP32, FP16, vanilla BFP, and FP8, focusing on the tradeoffs between accuracy and precision. 
To measure the improvement DBFP brings to the hardware, we compare its hardware metrics with SOTA Softmax acceleration architectures: Hyft \cite{xia2023softmax}, TCAS-I’22 \cite{zhang2023high},  ISCAS'23 \cite{koca2023hardware} and TCAS-II’22 \cite{s2021reducedsoftmaxunitdeep}.

\noindent \textbf{Models and Datasets.}
We evaluate DB-Attn on both LLM and Vision models. For LLMs, we test on LLaMA-7B, LLaMA2-7B, and LLaMA3-8B \cite{touvron2023llama, touvron2023llama2, meta2024llama3}, using WikiText2 and C4 datasets for perplexity comparison. Zero-shot accuracy is evaluated on PIQA, ARC, BoolQ, HellaSwag, and Winogrande tasks. For Vision tasks, we test image classification using ViT and Swin \cite{dosovitskiy2020image, liu2021swin} on ImageNet, and object detection using Detr on COCO dataset.
\noindent \textbf{Implementations.}
We implement DB-Attn on NVIDIA A6000 GPU using Pytorch and huggingface, replacing only the Attention layer with DB-Attn. We use 128-element blocks along matrix rows, with 8-bit mantissa and 5-bit shared exponent for both DBFP and BFP formats. DBFP is applied to Softmax operations and matrix multiplications in Attention Modules, using 7-bit DH-LUT for Softmax.

We implement DBFP in RTL using Chisel \cite{bachrach2012chisel}, verify with Verilator, and deploy on AMD Alveo™ U280 FPGA using Vivado. We compare our design with other FPGA-based Softmax accelerators. For accurate area and power analysis, we synthesize the accelerator using Synopsys Design Compiler at 2.0 GHz on 28nm TSMC process.

\begin{figure*}[t]
\includegraphics[width=1\textwidth]{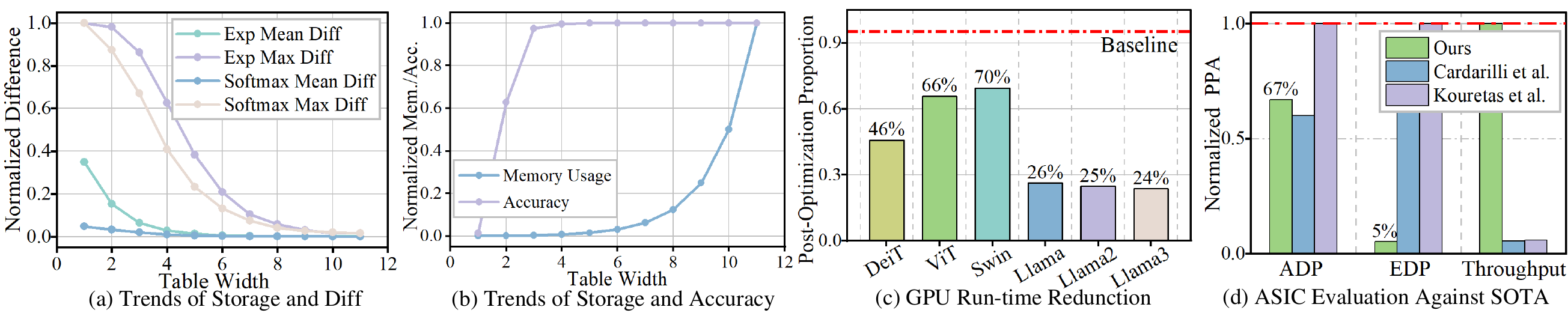}
\caption{Pipeline's balanced proportion under input sequences length growth.}\label{fig:fig5}
\end{figure*}


\subsection{Accuracy Results of DB-Attn}

\subsubsection{LLM Tasks.}
We examine the accuracy of DB-Attn on the language generation task and six zero-shot tasks on LLaMA LLMs, comparing it against vanilla BFP and FP8 format. Tab.~\ref{table:table1} presents the perplexity and zero-shot accuracy of LLMs. 
Wherein FP8 e4m3-S denotes the use of scaling factor  (the maximum value that FP8 e4m3 can represent) to rescale the value within the representable range of FP8 e4m3. 
It can be seen in Tab.~\ref{table:table1} that the direct application of vanilla BFP format to Softmax operations leads to substantial accuracy drop, with LLaMA3's average accuracy on zero-shot tasks decreasing by 33.81\%. 
Similarly, FP8 e4m3 is inadequate for Attention calculations due to its inability to represent infinity.
DB-Attn outperforms both vanilla BFP and FP8 formats across all evaluated tasks, maintaining nearly the same accuracy as floating-point. 
These results unequivocally demonstrate the comprehensive excellence of DB-Attn in maintaining model performance while potentially reducing computational overhead.

\subsubsection{Vision Tasks.}
We assess tasks of object detection and image classification (Tab. \ref{table:table2}).
It is seen that DB-Attn's performance is similar to results on LLMs. 
DB-Attn can be losslessly integrated into existing Transformer models, showing its generalization and versatility across various distributions.

\subsubsection{Precision-to-Accuracy Pareto Frontier.} 
We test the computational error, LUT memory usage, and actual model accuracy of Softmax in DB-Attn under different LUT bit widths. To find the Pareto frontier of the LUT bit width configuration, we visualize some results in Fig.~\ref{fig:fig5}(a) and (b) -- when the LUT bit width is 5-7, a balance is achieved among computational error, memory usage, and accuracy.

\subsection{Hardware Implement Evaluation}

As BFP has been validated on linear operations in previous work \cite{yeh2022like, zou2024bie}, 
we mainly focus on the performance of hardware deployments for the yet to be well optimized nonlinear operation  Softmax.

\subsubsection{DBFP GPU Run-time Analysis.} 
We implement a custom CUDA Softmax operator to emulate DBFP formats on NVIDIA A800 GPU. This operator replaces the Softmax in various models (e.g., LLaMA and ViT) and performs inference.
Results in Fig.~\ref{fig:fig5}(c)  demonstrate that DBFP-based Softmax consistently achieves speed improvements of at least 30\% across diverse model architectures. Notably, on the LLaMA series, we reduced latency by 74\% on average.

\subsubsection{DBFP Hardware Implement on FPGA.}

We evaluate designs against SOTA based on Softmax processing latency, FPGA resource utilization (LUT and FF), maximum operating frequency, and FOM (a comprehensive metric).

\begin{small}
\begin{equation}\label{eq:eq11}
\text{FOM}=\frac{F_{\text{max}} \times N \times W}{\text{LUT}+\text{FF}}
\end{equation}
\end{small}
, where W and N denote the precision and numbers of the inputs. A higher FOM value indicates better performance. 
\begin{table}
\centering
\renewcommand\arraystretch{1.25}
\resizebox{1.0\linewidth}{!}{
\begin{tabular}{cccccccc}
\hline
\toprule[2pt]
Methods           & NUM            & Format   &  LUT & FF & \begin{tabular}[c]{@{}c@{}}Fmax\\ (MHz)\end{tabular} & \begin{tabular}[c]{@{}c@{}}Latency\\ (ns)\end{tabular} & FOM    \\ 
\midrule[1.5pt]
Xilinx FP        & 8            & FP32     & 13254 & 18664    & 435   & 232.3  & 3.488  \\ 
Hyft16           & 8            & FP16     & 1072  & 824       & 625   & 12.4   & 42.194 \\ 
Hyft32           & 8            & FP32     & 2399  & 1528      & 526   & 19     & 34.290 \\ 
TCAS-I’22         & 10           & Fixed 16 & 1476  & 698       & 500   & NA     & 36.798 \\ 
ISCAS’23          & 8            & FP16     & 909   & 333        & 476   & 10.5   & 49.056 \\ 
TCAS-II’22        & 1            & FP16     & 128   & 97         & 588   & 22.1     & 41.813 \\ 
\cellcolor[rgb]{ .906,  .902,  .902}\textbf{Ours}     &\cellcolor[rgb]{ .906,  .902,  .902}\cellcolor[rgb]{ .906,  .902,  .902}\textbf{1024} & \cellcolor[rgb]{ .906,  .902,  .902}\textbf{DBFP}    & \cellcolor[rgb]{ .906,  .902,  .902}\textbf{10872}  & \cellcolor[rgb]{ .906,  .902,  .902}\textbf{3743}   & \cellcolor[rgb]{ .906,  .902,  .902}\textbf{455}   & \cellcolor[rgb]{ .906,  .902,  .902}\textbf{73}  & \cellcolor[rgb]{ .906,  .902,  .902}\textbf{509.563} \\ 
\bottomrule[2pt]
\end{tabular}
}
\caption{SOTA Softmax accelerators comparison on FPGAs}
\label{table:table3}
\end{table} 


Tab. \ref{table:table3} shows our comparison with multiple SOTA designs, including Xilinx FP IP \cite{koca2023hardware} baseline. While existing Softmax accelerators only support input bandwidths under 16, our design uniquely accommodates the larger bandwidths needed for modern LLMs. Testing with 1024-length sequences, our implementation achieves 54.21\% less resource usage while operating at higher frequencies, reduces processing latency by 62.5\%, and delivers 128x higher computational bandwidth. Notably, it shows a 10x FOM improvement over ISCAS'23 SOTA, demonstrating the significant potential of our approach for nonlinear operation hardware units.

\subsubsection{Design Evaluation on ASIC.}
ASIC implementations can be supplemented with more accurate data on power consumption, maximum clock frequency, and scalability for high-volume applications. 
We evaluate the hardware design on ASIC using four key metrics: Area-Delay-Product (ADP, Area $\times$ Latency), Energy-Delay-Product (EDP, Energy $\times$ Latency), and Throughput (Freq $\times$ Bandwidth).

For scenarios with a uniform input sequence length of 1024, we normalized the experimental results to 28nm. 
Fig.~\ref{fig:fig5} (d) compares our design's normalized PPA (Power, Performance, and Area) 
 metrics with SOTA designs \cite{cardarilli2021pseudo, kouretas2020hardware}. 
Our design is capable of handling large bandwidth and long input sequences.
Hence, there is an average 10\% increase in area compared to the SOTA design.
However, this increase is offset by a significant enhancement in both energy consumption and throughput, exceeding 10x.  
In this context, the additional area requirement is considered a tolerable trade-off for substantial improvements in performance and efficiency.

\begin{figure}
\includegraphics[width=0.47\textwidth]{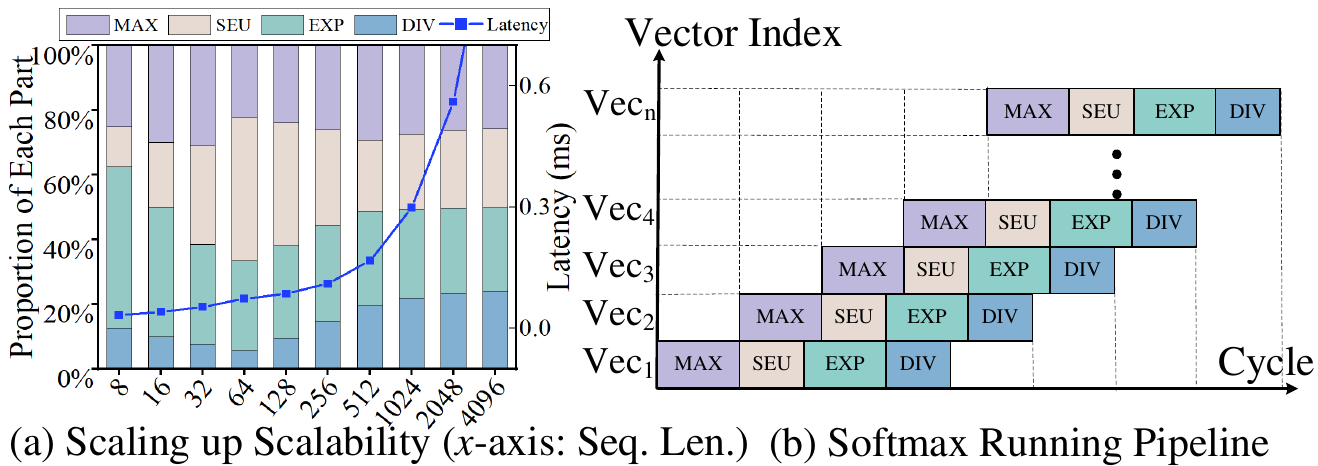}
\caption{Pipeline's balance under sequences scaling up.}\label{fig:fig6}
\end{figure}

\subsubsection{Hardware Scalability.}
The design keeps scalability in mind.
To demonstrate this feature, we conducted tuning tests with input sizes ranging from 8 to 4096 elements. Fig.~\ref{fig:fig6} shows the total latency and each part of the processing pipeline.
The total computation time grows exponentially with input size due to the quadratic relationship between input length and the size of the processed matrix. 
The scaled histogram in Fig.~\ref{fig:fig6} shows the balanced growth in the proportion of time consumed by each pipeline level as the input size increases.
No single component shows disproportionate latency growth; instead, pipeline allocation becomes more balanced with longer sequences.
These observations evidenced the parallelism and scalability of the design.

\section{Conclusion}
We present DBFP, an enhanced BFP variant optimizing nonlinear operations. Our DB-Attn framework, enables efficient Attention, advancing narrow-precision LLM inference.

\noindent \textbf{Lessons learned.}
Different from the conventional solutions that attempted to optimize nonlinear computation solely through hardware design or software techniques, this work shows that using an algorithm/hardware co-designed approach (DB-Attn), computation latency and memory accesses can be largely improved with light overhead.
The construction of DBFP and the algorithm-driven hardware provide key insights and effective means that foster a collaborative
environment of hardware and the algorithm for LLMs research that neither discipline could achieve independently.


\section{Acknowledgments}
The authors would like to thank the anonymous reviewers for their insightful and helpful feedback. 
This work is supported by the National Key Research and Development Program (Grant No.2024YFB4405600), the National Natural Science Foundation of China (Grant No. 62472086), the Basic Research Program of Jiangsu (Grants No. BK20243042 and NO. BG2024010), and the Start-up Research Fund of Southeast University (Grant No. RF1028624005).

\bibliography{aaai25}

\end{document}